\documentclass[prl,twocolumn,superscriptaddress,showpacs,amsmath]{revtex4}
\usepackage{graphicx}

\begin{document}

\def\ket#1{|#1\rangle}
\def\bra#1{\langle#1|}
\def\av#1{\langle#1\rangle}
\def\myarrow{\mathop{\longrightarrow}}

\title{Minimum-energy pulses for quantum logic cannot be shared}

\author{J. Gea-Banacloche}
\email[]{jgeabana@uark.edu}
\affiliation{Department of Physics, University of Arkansas, Fayetteville, AR 72701}
\author{Masanao Ozawa}
\email[]{ozawa@math.is.tohoku.ac.jp}
\affiliation{Graduate School of Information Sciences, Tohoku University, Aoba-ku, Sendai 980-8579, Japan}

\date{\today}

\begin{abstract}
We show that if an electromagnetic energy pulse with average photon number $\bar n$  is used to carry out the same quantum logical operation on a set of $N$ atoms, either simultaneously or sequentially, the overall error probability in the worst case scenario (i.e., maximized over all the possible initial atomic states) scales as $N^2/\bar n$.  This means that in order to keep the error probability bounded by $N\epsilon$, with $\epsilon \sim 1/\bar n$, one needs to use $N\bar n$ photons, or equivalently $N$ separate ``minimum-energy'' pulses: in this sense the pulses cannot, in general, be shared.  The origin for this phenomenon is found in atom-field entanglement.  These results may have important consequences for quantum logic and, in particular, for large-scale quantum computation.
\end{abstract}


\maketitle

There is by now a substantial amount of evidence \cite{ozawa1,jgb3} that an elementary quantum logical operation on a qubit requires a minimum amount of energy which is inversely proportional to the acceptable error probability $\epsilon$.  This has been most extensively studied for atomic systems interacting with an electromagnetic-pulse control field \cite{barnes,vanenk,jgb1,geaozawa} with the generic result that the error probability scales as the inverse of the number of photons in the (quantized) pulse.  

A question that has not so far been addressed is whether this minimum energy must truly be dedicated to each qubit and each operation, or whether it could be shared by two or more qubits on which one wanted to perform a given operation, either sequentially or simultaneously.  Intuitively, one would expect the latter to be the case: if a pulse of light containing, say, $10^5$ photons has just interacted with an atom that may at most add or subtract one photon to the field, one would not expect this very small change to make a substantial difference if the same pulse is used later to act on another atom. Also, it is a fact (and this point will be elaborated on later) that, for an atom or ion in free space, the most important consequence of field quantization is spontaneous emission \cite{itano,vanenk2,jgb2}; from this perspective, all that should matter is to have a sufficiently large electric field at the location of the atom, so as to complete the operation before it can decay, and there appears to be no reason why two or more atoms should not be able to share this field, for a sufficiently long or wide pulse, without an appreciable increase in the error rate.

In contrast to these very reasonable expectations, we show here that a minimum energy pulse \emph{cannot}, in general, be shared as described above: specifically, the result to be proven is that if the same pulse, with average photon number $\bar n$, is used to carry out the same quantum logical operation on a set of $N$ identical atoms, either simultaneously or sequentially, the overall error probability in the worst case scenario (i.e., maximized over all the possible initial atomic states) scales, not as $N/\bar n$, as one would expect for $N$ independent processes, but as $N^2/\bar n$.  This means that in order to keep the maximum error probability bounded by $N\epsilon$, with $\epsilon \sim 1/\bar n$, one needs to use a total of $N/\epsilon$ photons, that is to say, the energy equivalent to $N$ separate ``minimum energy pulses.''

The most general proof of this result makes use of the methods of \cite{geaozawa,ozawa2} and applies to a system of $N$ atoms interacting with an arbitrary set of quantized field modes via a Hamiltonian of the form
\begin{align}
H = &\hbar\sum_{\bf k}\omega_{\bf k} a_{\bf k}^\dagger a_{\bf k} +\frac{\hbar\omega_0}{2}\sum_{i=1}^N \sigma_{iz} \notag \\
&+i\hbar\sum_{{\bf k}, i} g_{\bf k} \left(U_{{\bf k},i} a_{\bf k} \sigma_i^\dagger - U_{{\bf k},i}^\ast a_{\bf k}^\dagger \sigma_i \right)
\label{e1}
\end{align}
Here the $g_{\bf k}$ are coupling constants and the $U_{{\bf k},i}$ are arbitrary mode functions, evaluated at the positions (indexed by $i$) of the respective atoms.  We use the convention $\sigma_{iz}\ket e_i = \ket e_i$, where $\ket e_i \equiv \ket 0_i$ is the excited state of the $i$-th two-level atom.  The model (\ref{e1}) is extremely general, and it can easily be further generalized to cover multilevel atoms and Raman-type processes (see \cite{geaozawa} for details); in particular, it includes spontaneous emission implicitly, by the presence of quantized vacuum modes.

The key property of the Hamiltonian (\ref{e1}) is that it has a conserved quantity, namely
\begin{equation}
L= L_1 + L_2 = \frac{1}{2}\sum_{i=1}^N \sigma_{iz} + \sum_{\bf k} a_{\bf k}^\dagger a_{\bf k}
\label{e2}
\end{equation}
where $L_1$ is an atomic operator and $L_2$ a field operator.  Suppose we want to use the Hamiltonian (\ref{e1}) to implement a certain quantum logical operation so that, after a time $T$ (omitted below for simplicity) the evolution operator $U$ is as close as possible to some desired $U_\text{ideal}$.  We can get an idea of how successful the procedure is by looking at how an atomic operator $A$ is transformed, and specifically at the difference $D\equiv U^\dagger A U - U_\text{ideal}^\dagger A U_\text{ideal}$.  If we choose $A$ so that it commutes with $L_1$, the methods of \cite{geaozawa} can be used to show that one must have $\av{D^2} \ge \sigma(D)^2 \ge |\av{[L_1,U^\dagger_\text{ideal} A U_\text{ideal}]}|^2/\sigma(L)^2$, where $\sigma(\cdot)$ stands for the standard deviation of an operator, and all expectation values are calculated in the initial state, which we shall take to be of the form $\ket\psi\ket\Phi$, with $\ket\psi$ an atomic state and $\ket\Phi$ a field state.  

Consider the case in which $U_\text{ideal}$ corresponds to a collective $\pi/2$ pulse, which is a Hadamard gate up to an overall bit flip.  Specifically,
\begin{equation}
U_\text{ideal} = \frac{1}{2^{N/2}}\begin{pmatrix}1 & -1 \\ 1 & 1\end{pmatrix}^{\otimes N}
\label{e4}
\end{equation}
Then choosing $A=\prod_{i=1}^N \sigma_{iz}$, and the initial atomic state $\ket{\psi} = (\ket{0}^{\otimes N} + i \ket{1}^{\otimes N})/\sqrt 2$, one immediately obtains
\begin{equation}
\av{D^2} \ge \frac{N^2}{2N(N+1) + 4\sigma\left(\sum_{\bf k} n_{\bf k}\right)^2}
\label{e5}
\end{equation}
If the initial field state is a multimode coherent state, one has $\sigma(\sum_{\bf k} n_{\bf k})^2 = \sum_{\bf k} \av{n_{\bf k}}=\bar n$, and Eq.~(\ref{e5}) exhibits the $N^2/\bar n$ scaling, as long as $\bar n \gg N^2$.  

To relate the error $\av{D^2}$ in the operator $A$ to a more familiar error measure, such as the overall fidelity, one can follow a procedure similar to the one in Appendix A of \cite{geaozawa}.   With the above choices for $A$ and $\ket\psi$ 
\begin{align}
&\av{D^2} = \notag \\
&\frac 1 2 \left\| \left(\prod\sigma_{iz} U - (-1)^N U \prod\sigma_{ix}\right)\left(\ket{0}^{\otimes N} + i \ket{1}^{\otimes N}\right)\ket{\Phi}\right\|^2
\label{e6}
\end{align}
For definiteness, assume that $N$ is even. Inserting a sum $\sum\ket{\psi'}\bra{\psi'}$ over a complete set of atomic states, and noting that $\bra{\psi'}\prod\sigma_{iz} = (-1)^p\bra{\psi'}$, where $p$ is the number of ones in $\ket{\psi'}$, one obtains
\begin{align}
\av{D^2} = & \sum_{\text{$p$ even}}\left\|\bra{\psi'}U((\ket{0}^{\otimes N} -  \ket{1}^{\otimes N})\ket{\Phi}\right\|^2 \notag \\
& + \sum_{\text{$p$ odd}}\left\|\bra{\psi'}U((\ket{0}^{\otimes N} +  \ket{1}^{\otimes N})\ket{\Phi}\right\|^2
\label{e7}
\end{align}
However, with $N$ even, the ideal operation (\ref{e4}) when acting on the state 
$((\ket{0}^{\otimes N} -  \ket{1}^{\otimes N})/\sqrt 2$ produces a superposition of states with only odd numbers of ones, which means that the first term in (\ref{e7}) is $\le 2(1-{\cal F}_-^2)$, where ${\cal F}_-^2$ is the fidelity for that initial state.  Similarly, the second term is $\le 2(1-{\cal F}_+^2)$, where ${\cal F}_+^2$ is the fidelity for the initial state $((\ket{0}^{\otimes N} + \ket{1}^{\otimes N})/\sqrt 2$.  Both of these infidelities must be smaller than the infidelity $(1-{\cal F}^2)_\text{max}$, maximized over all initial atomic states, and therefore one concludes
\begin{equation}
(1-{\cal F}^2)_\text{max} \ge \frac{1}{4}\frac{N^2}{2N(N+1) + 4\bar n}
\label{e8}
\end{equation}
The same result for $N$ odd can be established along similar lines.

While the above method is very powerful, it does not provide much insight on the origin of the $N^2$ scaling, and it has also proven hard to generalize it to other operations such as bit flips ($\pi$ pulses). Hence it is worthwhile to explore a much simpler model for the interaction which was shown in \cite{geaozawa} to capture the essence of the constraints arising from the conservation of (\ref{e2}).  This is a single-mode model where the creation and annihilation operators are replaced by $e^{\pm i\hat\phi}$, where $\hat\phi$ is a ``phase operator'':
\begin{equation}
H=i\hbar\Omega\left(e^{i\hat\phi}\sum_{i=1}^N \sigma_{iz}^\dagger - e^{-i\hat\phi}\sum_{i=1}^N \sigma_{iz}\right)
\label{e9}
\end{equation}
Although a Hermitian phase operator, strictly speaking, does not exist in the full Fock space, reasonable approximations can be defined \cite{barnett} with the desired properties, namely, $e^{\pm i\hat\phi}\ket n = \ket{n\mp 1}$.  As also shown in \cite{geaozawa}, the following manipulations will be accurate enough provided the weight of the vacuum in the initial field state $\ket\Phi$ is vanishingly small, which is always the case for a coherent state with a high excitation number.  The model (\ref{e9}) removes the nonessential (in this context) complication of the field amplitude fluctuations, and captures the basic requirement expressed by the conservation of (\ref{e2}), namely, that the photon number must increase or decrease by 1 when any of the atoms makes a transition.

Integration of (\ref{e9}) is trivial.  Assuming each atom interacts with the field for a total time $T$ (it does not matter whether simultaneously or sequentially), the evolution operator is
\begin{equation}
U=\begin{pmatrix}\cos\Omega T & -e^{-i\hat\phi}\sin\Omega T \\ e^{i\hat\phi}\sin\Omega T & \cos\Omega T\end{pmatrix}^{\otimes N}
\label{e10}
\end{equation}
When $\Omega T = \pi/4$ (the $\pi/2$-pulse condition) (\ref{e10}) would reduce to (\ref{e4}) provided $\hat\phi =0$.  In what follows, it will be assumed that $\av{\hat\phi}=0$ and $\av{{\hat\phi}^2}\equiv \sigma(\hat\phi)^2$ is small. Taking again the initial atomic state to be of the form $(\ket{0}^{\otimes N} + i \ket{1}^{\otimes N})/\sqrt 2$, which would be transformed by $U_\text{ideal}$ into $(\ket{+x}^{\otimes N} + (-1)^N \ket{-x}^{\otimes N})/\sqrt 2$, a direct calculation using (\ref{e10}) yields the fidelity
\begin{align}
{\cal F}^2 &= \frac{1}{2^{2(N+1)}}\Bigl\|\bigl[(1+e^{i\hat\phi})^N+(e^{-i\hat\phi}+1)^N 
\notag \\
&\qquad+ i (1-e^{-i\hat\phi})^N - i (e^{-i\hat\phi}-1)^N \bigr]\ket\Phi \Bigr\|^2 \notag \\
&\simeq 1 - \frac{N(N+1)}{4}\av{{\hat\phi}^2}
\label{e11}
\end{align}
For a coherent state we have $\sigma(\phi)^2 \simeq 1/4\bar n$, and hence the result
\begin{equation}
(1-{\cal F}^2)_\text{max} \ge \frac{N(N+1)}{16\bar n} \qquad \text{($\pi/2$ pulse)}
\label{e12}
\end{equation}
The right-hand side of (\ref{e12}) is always greater than that of (\ref{e8}), as it should, and, in fact, the two expressions agree up to terms of the order of $(N^2/\bar n)^2$.

In spite of its simplicity, there are several reasons to expect that the single-mode approach provides a universally valid lower bound to the infidelity.  First, because adding more modes generally only adds more avenues for decoherence (a point that will be elaborated on later), and second, because it has been shown by Silberfarb and Deutsch \cite{silberfarb} that the atom-field entanglement predicted by single-mode models (specifically, the Jaynes-Cummings model) is actually a good approximation to the actual entanglement obtained from multimode, free-space calculations, as long as the total probability for spontaneous emission over the duration of the gate remains small (which is the regime in which one would want to operate in any case).

With this in mind, one can use the model (\ref{e9}) to show that the $N^2/\bar n$ scaling also applies to the case of bit flips (or $\pi$ pulses).  This is obtained by setting $\Omega T = \pi/2$ in Eq.~(\ref{e10}).  Again starting from a state of the GHZ form, $(\ket{0}^{\otimes N} + e^{i\delta} \ket{1}^{\otimes N})/\sqrt 2$ (with arbitrary phase $\delta$), one finds for the fidelity ${\cal F}^2 = \av{\cos^2(N\hat\phi)}$, and therefore, in a coherent state
\begin{equation}
(1-{\cal F}^2)_\text{max} \ge \frac{N^2}{4\bar n} \qquad \text{($\pi$ pulse)}
\label{e13}
\end{equation}

The bit-flip example is especially helpful to show how the effect described arises from atom-field entanglement.  In an $N$-atom bit flip, the initial states $\ket 0^{\otimes N}\ket\Phi$ and $\ket 1^{\otimes N}\ket\Phi$ would have to become $\ket 1^{\otimes N}a^N\ket\Phi/{\cal N}_1$ and $\ket 0^{\otimes N}{a^\dagger}^N\ket\Phi/{\cal N}_2$, respectively (where ${\cal N}_1$ and ${\cal N}_2$ are appropriate normalization constants), and therefore the coherent superposition $(\ket 0^{\otimes N}+\ket 1^{\otimes N})/\sqrt 2$, which ideally should be left invariant by the operation, is instead transformed into
\begin{equation}
\frac{1}{\sqrt 2}\left(\ket 0^{\otimes N}\frac{{a^\dagger}^N\ket\Phi}{{\cal N}_2}+\ket 1^{\otimes N}\frac{a^N\ket\Phi}{{\cal N}_1}\right)
\label{e14}
\end{equation}
This superposition differs from the intended result because the atomic and field states are entangled, since the field states $a^N\ket\Phi/{\cal N}_1$ and ${a^\dagger}^N\ket\Phi/{\cal N}_2$ are different in general.  In fact, the infidelity of the state (\ref{e14}) is simply proportional to the ``lack of overlap'' between the two field states: 
\begin{equation}
1-{\cal F}^2 = \frac 1 2 - \frac{1}{4{\cal N}_1{\cal N}_2}\left(\bra\Phi a^{2N}\ket\Phi + \bra\Phi {a^\dagger}^{2N}\ket\Phi \right)
\label{e15}
\end{equation}
Now, one might think that for a very ``classical'' state $\ket\Phi$, with a large average photon number, the difference between the state resulting from the creation of $N$ photons and the one resulting from the annihilation of $N$ photons would be very small, and it is---but, somewhat surprisingly, it turns out to be quadratic, rather than linear, in $N$.  Specifically, for a coherent state $\ket\alpha$, with $|\alpha|^2=\bar n$, the only nontrivial expectation value appearing in (\ref{e15}) is ${\cal N}_2 = (\bra\alpha a^N {a^\dagger}^N\ket\alpha)^{1/2}$, for which one has (see \cite{agarwal})
\begin{equation}
{\cal N}_2^2 = \sum_{n=0}^N \frac{{\bar n}^n (N!)^2}{(n!)^2 (N-n)!} 
= {\bar n}^N\left[1+\frac{N^2}{\bar n} + O\left(\left(\frac{N^2}{\bar n}\right)^2\right) \right]
\label{e16}
\end{equation}
and using this in (\ref{e15}) one obtains $1-{\cal F}^2 = {N^2}/{4\bar n} + O(({N^2}/{\bar n})^2)$ in agreement with (\ref{e13}).  

%Incidentally, this agreement shows that nothing essential was lost by using the simpler raising and lowering operators $e^{\pm i\hat\phi}$ in (\ref{e9}) in place of the creation and annihilation operators $a$ and $a^\dagger$.

This derivation suggests the kinds of situations when one may expect the $N^2$ terms in the infidelity to be significant: when (as in (\ref{e14})) the final state wavefunction contains at least two terms, with reasonably large weights, that differ from each other by the action of a number of creation operators of the order of $N$.

At this point it may be thought that a way to avoid this kind of difficulty in quantum logical operations would be to use an encoding such as $\ket{0}_L = \ket{01}$, $\ket{1}_L = \ket{10}$ \cite{lidar,ozawa3}, where each logical qubit is represented by two physical qubits, and the numbers of ones and zeros in the states $\ket{0}_L$ and $\ket{1}_L$ are the same. It is also known that such an encoding makes the logical qubit insensitive to collective phase fluctuations, such as those in Eq.~(\ref{e9}) \cite{kielpinski}.  However, since the Hamiltonian (\ref{e1}) does not couple directly the states $\ket{01}$ and $\ket{10}$, a meaningful discussion of what can or cannot be done with encoded qubits requires a careful look at the ``effective Hamiltonians'' that describe the action of the control fields on the encoded states.  For example, in the proposal \cite{kielpinski} to use the above encoding in an ion trap, in conjunction with the S\o rensen-M\o lmer gate \cite{sorensen}, one obtains, in effect, an evolution operator of the form $U=\cos(g \hat n t) + i \sin(g \hat nt) \sigma_X$, where $\sigma_X$ is the encoded bit-flip operator, and $\hat n$ is a photon number operator (or a sum of such operators).  But then one can show explicitly that the $N^2/\bar n$ scaling must hold, for certain initial states, for arbitrary operations.  

To exhibit this for a collective bit-flip, let $\ket{\pm X}$ be the (two-qubit) encoded eigenstates of $\sigma_X$.  Then $U\ket{\pm X}\ket{\Phi} = e^{\pm ig\hat n t}\ket{\pm X}\ket{\Phi}$.  Separating $\hat n$ into average $\bar n$ and fluctuations $\Delta \hat n$, where, for a bit flip operation, $g\bar n T = \pi/2$, and making  $U^{\otimes N}$ act on a superposition $2^{-1/2}(\ket{+ X}^{\otimes N}+ \ket{-X}^{\otimes N})\ket{\Phi}$, the result is (up to a global phase)
\begin{equation}
\frac{1}{\sqrt 2}\left(e^{igN\Delta \hat n T} \ket{+ X}^{\otimes N} + (-1)^Ne^{- igN\Delta \hat n T} \ket{- X}^{\otimes N}\right) \ket{\Phi}
\label{e16b}
\end{equation}
This is to be compared to the action of $U^{\otimes N}_\text{ideal} \equiv i^N\sigma_X^{\otimes N}$, which yields the same state except for the $e^{\pm igN\Delta \hat n T}$ terms. The infidelity is then easily calculated to be $1-{\cal F}^2 = \av{\sin^2(g N \Delta \hat n T)} \simeq \pi^2 N^2 \sigma(n)^2/4{\bar n}^2 = \pi^2 N^2 /4{\bar n}$ for a coherent state.  

The above derivation, concerning what is arguably the most popular proposal for encoded logic in an atomic system, is enough to make us skeptical that one might get around the $N^2/\bar n$ scaling using these approaches.  Nonetheless, other encodings and gate mechanisms certainly exist, and we do intend to look into as many of them as possible in the future.

Finally, we would like to supplement our single-mode calculations by considering briefly their possible relevance for a collection of atoms or ions in free space, where spontaneous emission is the leading quantized-field source of decoherence.  Let the laser beam (possibly including refocusing in between atoms) be taken to define an effective single mode, all the other modes being then in the vacuum state.  Every atom has a probability $p$ to emit a photon in the course of the interaction, and if the beam waist at the location of the atom is $w_0$ then the probability that the photon goes into the laser mode is of the order of $3\lambda^2/8\pi^2w_0^2 \equiv \lambda^2/A$, where the area $A$ is of the order of the cross-section of the beam (see Eq.~(1) of \cite{jgb2}).  Adopt a simple model in which a photon being emitted outside of the laser mode, by any atom, leads to the total failure of the operation.  The overall failure probability of, e.g., a collective $\pi$ pulse can then be written (assuming $pN\ll 1$) as
\begin{align}
P_f &\simeq Np\left(1-\frac{\lambda^2}{A}\right) + \left[1-p\left(1-\frac{\lambda^2}{A}\right)\right]^N\frac{N^2}{4 \bar n}\notag \\
&\simeq Np\left(1-\frac{\lambda^2}{A}\right) + \frac{N^2}{4 \bar n}
\label{e17}
\end{align}
where the second term accounts for the result of the single-mode analysis given above (Eq.~(\ref{e13})), in the case that all the photons are emitted into the laser pulse, with $\bar n$ being the number of photons in the pulse. Now, it was shown in \cite{jgb2} that for an operation such as a $\pi$ or $\pi/2$ pulse, on resonance, $p$ was of the order of $1/\bar n$ times a geometric factor of the order of $2\pi^4w_0^2/3\lambda^2 = \pi^2 A/4\lambda^2$, and so $P_f$ becomes
\begin{equation}
P_f \sim \frac{N}{4\bar n}\left(\frac{\pi^2 A}{\lambda^2} + N\right)
\label{e18}
\end{equation}
This exhibits a scaling that is quadratic in $N$ for sufficiently large $N$, but is only linear in $N$ if $\pi^2A/\lambda^2 \gg N$,  However, the reason for this apparent ``linearity'' is that in this case, because of the suboptimal coupling between the atoms and the field, one is already using many more photons than one would have to in the optimal, single-mode case.  Indeed, in the single-mode treatment, the requirement to keep the overall failure probability smaller than $N\epsilon$ is $\bar n > N/4\epsilon$, whereas from (\ref{e18}) one requires $\bar n > (\pi^2 A/4 \lambda^2)/\epsilon$, which is $\gg  N/4\epsilon$ in this limit.  
%This simply shows the general principle that it should not be possible to lower one's energy requirements by reducing the effective atom-field coupling.  

In conclusion, we have shown that, in general, the minimum field energy needed to carry out a quantum logical operation on a set of $N$ atomic qubits with a given overall error probability scales as $N^2$; or, equivalently, that in order to ensure a constant error rate per operation per qubit, (say, $\epsilon$) one needs to use, at least, the total energy of $N$ ``minimum energy'' pulses, with $\bar n \sim 1/\epsilon$.  In this sense, minimum energy pulses cannot be shared.

%The $N^2$ scaling has been shown to arise from atom-field entanglement, together with the somewhat surprising fact that, for a coherent state, the addition of $N$ photons leads to a ``lack of overlap'' with the initial state which is quadratic in $N$.  This means that the quantum nature of the fields used in quantum logical operations may become important much earlier than one would naively expect, since the number to be compared to the average photon number $\bar n$ may not be the total number of photons exchanged, $N$, but rather its \emph{square}.

Overall, these results will need to be taken into consideration when designing large-scale quantum computing devices, especially in the proposals that would rely on the simultaneous manipulation of many atoms by a single electromagnetic pulse.  Examples might include schemes for cluster state computation \cite{raussendorf}, and/or for quantum computing with atoms in optical lattices \cite{kay}.  (Note also that the results presented here are not restricted to atomic systems; they would apply equally well to, e.g., superconducting qubits manipulated by radiofrequency pulses.)  ``Bang-bang'' schemes for decoherence suppression \cite{viola} may also envision the simultaneous flipping of many qubits by a single pulse; the results presented here clearly place a constraint on the minimum energy required to carry out such operations with an acceptable error rate.

\begin{acknowledgments}
This research has been supported by the National Science Foundation and a Grant-in-Aid for Scientific Research of the JSPS.
\end{acknowledgments}

\end{document}